\documentclass[
preprintnumbers,
twocolumn,
nofootinbib,
amsmath,amssymb,
 aps,
]{revtex4-2}

\usepackage{graphicx}
\usepackage{dcolumn}
\usepackage{bm}
\usepackage{hyperref}

\usepackage{comment}
\usepackage{caption}
\usepackage{subcaption}
\usepackage{booktabs}
\numberwithin{equation}{section}
\usepackage{color}
\usepackage{mathtools}
\usepackage[normalem]{ulem} 

\usepackage{url}
\hypersetup{colorlinks=true, citecolor=cyan}

\newcommand{\be}{\begin{equation}}
\newcommand{\ee}{\end{equation}}

\begin{document}
\preprint{RESCEU-5/24}
\preprint{YITP-24-30}
\preprint{RIKEN-iTHEMS-Report-24}

\title{Slowly decaying ringdown of a rapidly spinning black hole II: \\Inferring the masses and spins of supermassive black holes with LISA}
\author{Daiki Watarai$^{1,2}$, Naritaka Oshita$^{3,4,5}$, and Daichi Tsuna$^{6,2}$}

\affiliation{${}^1$Graduate School of Science, The University of Tokyo, Tokyo 113-0033, Japan, \\${}^2$Research Center for the Early Universe (RESCEU), Graduate School of Science, The University of Tokyo, Tokyo 113-0033, Japan, \\${}^3$Center for Gravitational Physics and Quantum Information, Yukawa Institute for Theoretical Physics,
Kyoto University, 606-8502 Kyoto, Japan, \\${}^4$The Hakubi Center for Advanced Research, Kyoto University,
Yoshida Ushinomiyacho, Sakyo-ku, Kyoto 606-8501, Japan, \\${}^5$RIKEN iTHEMS, Japan, \\${}^6$TAPIR, Mailcode 350-17, California Institute of Technology, Pasadena, California 91125, USA}

\newcommand*{\diff}{\,\mathrm{d}}

\date{\today}

\begin{abstract}
Electromagnetic observations reveal that almost all galaxies have supermassive black holes (SMBHs) at their centers, but their properties, especially their spins, are not fully understood. Some of the authors have recently shown [Oshita and Tsuna (2023)] that rapid spins of $>0.9$, inferred for masses around $10^7\ M_\odot$ from observations of local SMBHs and cosmological simulations, source {\it long-lived} ringdowns that enhance the precision of black hole spectroscopy to test gravity in the near-extreme Kerr spacetime. In this work, we estimate the statistical errors in the SMBH mass-spin inference in anticipation of the LISA's detection of extreme mass-ratio mergers. We show that for rapidly spinning SMBHs, more precise mass and spin measurements are expected due to the excitations of higher angular modes. For a near-extremal SMBH of mass $10^7M_\odot$ merging with a smaller BH with mass ratio $10^{-3}$ at a luminosity distance of $\lesssim 10\:\mathrm{Gpc}$ (redshift $z \lesssim 1.37$), the measurement errors in the mass and spin of the SMBH would be $\sim 1\:\mathrm{\%}$ and $\sim 10^{-1}\:\mathrm{\%}$ respectively.
\end{abstract}

\maketitle

\section{Introduction}
Various electromagnetic observations suggest the existence of supermassive black holes (SMBHs) at galactic centers, with a broad range of masses of $10^{5}$ -- $10^{10} M_\odot$ (e.g.~\cite{Greene_2010, John+_2013, Reynolds_2021}).
In general relativity (GR), astrophysical BHs are characterized by two parameters only, mass $M$ and angular momentum $J$. Accurate estimates of the mass and spin of SMBHs are desired in several astrophysical contexts. For example, the masses and spins of SMBHs are important for the BH growth history over cosmic time, appearance of nuclear transients like tidal disruption events \cite{Gezari2021}, and strength of active galactic nuclei (AGN) feedback \cite{Fiacconi18,Sebastian_Springel_2019,Bollati23}.
While studies on SMBH spin evolution using cosmological simulations predict that BHs of masses around $10^{7}M_\odot$ typically have near-extremal spins~\cite{Dubois14,Sebastian_Springel_2019} (in agreement with analysis of local SMBHs~\cite{Reynolds_2021}, and possibly high-redshift analogs~\cite{Inayoshi2024}), the evolution of mass and spin should highly depend on the astrophysical and cosmological models employed. These motivate us to accurately infer the mass and spin of SMBHs, especially for rapidly spinning ones, by an independent way.

Another interest arises from ``BH spectroscopy" in gravitational-wave (GW) astronomy, which aims to test gravity in strong gravity regimes by extracting the complex frequencies of quasi-normal modes (QNMs) from GW ringdown emitted by a BH~\cite{Dreyer_2004}. QNM frequencies are discretized complex values and are unique for the remnant mass and spin by virtue of the black hole no-hair theorem. QNM frequencies $\omega = \omega_{lmn}$ are labeled by the multipole mode $l,m$ and the overtone number $n$. One can test GR by comparing those frequencies with the QNM frequencies predicted by the linear BH perturbation theory. The LIGO-Virgo-KAGRA (LVK) collaboration has so far reported 90 GW events from compact binary coalescences (e.g.~\cite{testing_GR_GWTC1, testing_GR_GW150914, testing_GR_GWTC2, testing_GR_GWTC3}), and various types of verifications of GR in the ringdown, including the test of the no-hair theorem, have been conducted 
(e.g.~\cite{NoHairTest_2019, BHAreaTheoremTest_2021, Capano_2020, Eliot_2021, Sizheng_2022, correia2024low}). These tests require at least two QNMs to be confidently detected. Roughly speaking, one mode is used to estimate the mass and spin of the remnant, and the other is used to check the consistency with the results of the first mode assuming GR. 

For a comparable mass-ratio binary BH merger which is the main target of the LVK collaboration, numerical relativity simulations indicate that the quadrupole mode $(l,m)=(2,2)$ dominates the GW signal with the remnant spin of $j \sim 0.7$, where $j\equiv cJ/GM^2$ is the dimensionless spin parameter with $c$ and $G$ respectively the speed of light and the gravitational constant. Therefore, most of the existing ringdown data analysis focused on the consistency between the longest-lived fundamental mode, $(l,m,n)=(2,2,0)$, and the first overtone, $(2,2,1)$, for comparable-mass binaries (e.g.~\cite{Isi:2022mhy, Carullo:2023gtf}).
Although the (tentative) detection of a higher multipole mode $(3,3,0)$ for asymmetric-mass binaries (e.g.~\cite{Abedi:2023kot, capano2023statistical}) was proposed, it is still controversial whether the subdominant mode has been detected or not. 

On the other hand, for mergers of much smaller mass ratios higher multipole modes can be excited with larger relative amplitudes. One of the authors has numerically found that for a BH with high spin ($j>0.9$), GWs induced in the ringdown stage can be dominated by overtones and higher angular modes~\cite{Oshita_2021, Oshita_2023}. Especially, in the case of the near-extremal Kerr BH which has $j=0.99$, these modes are excited with amplitudes comparable to or larger than that of the quadrupole mode. In addition, for larger spins, a ringdown signal is long-lived due to smaller decay rates, which is advantageous for BH spectroscopy as pointed out by some of the authors in Ref.~\cite{Oshita_Tsuna_2023}. For binary stellar-mass BHs in the LVK bands, the remnant BH would likely not reach such a near-extremal spin due to strong angular-momentum loss of the BH progenitor's core before BH formation \cite{Fuller:2019sxi}, as well as the loss of orbital angular momentum during the inspiral phase~\footnote{As far as we are aware, the largest remnant spin modelled for equal-mass mergers is $j = 0.9507$ (from \texttt{SXS:BBH:1124} in the SXS catalog~\cite{Boyle_2019}), which resulted from a merger of two BHs having aligned spins of both $j = 0.998$.}. Therefore, small mass-ratio mergers involving rapidly spinning SMBHs could be an excellent probe to test gravity in the near-extreme Kerr spacetime. 

The Laser Interferometer Space Antenna (LISA)~\cite{Amaro-Seoane_2017, Amaro-Seoane_2023}, targeting detection of GWs of mHz frequencies, is planned to be launched in the next decade. The frequencies correspond to the QNM frequencies of SMBHs with masses around $10^{7}M_\odot$, which is right in the mass range where they are expected to have rapid spins of $j\gtrsim 0.9$~\cite{Reynolds_2021,Sebastian_Springel_2019}. Motivated by this, Ref.~\cite{Oshita_Tsuna_2023} discussed the feasibility of BH spectroscopy with LISA, where they focused on a compact object plunging into an SMBH in the intermediate or extreme mass-ratio regimes. They show that one can perform high-precision BH spectroscopy for rapidly spinning SMBHs. However, in their study the mass and spin of the SMBHs were assumed to be fully known, and the feasibility of LISA in constraining these parameters were not considered.

In this study, we significantly expand the analysis of Ref.~\cite{Oshita_Tsuna_2023}, and newly investigate how precisely the mass and spin of an SMBH can be inferred (assuming GR) from detecting intermediate or extreme mass ratio binary BH mergers with LISA. While several studies have discussed the feasibility of BH spectroscopy with LISA for SMBH mergers (e.g.~\cite{Berti_Cardoso_Will_2006, Baibhav_2019, Baibhav_2020, Bhagwat_2022}), we concentrate on rapidly spinning SMBHs as was investigated in Ref.~\cite{Oshita_Tsuna_2023}.

This paper is organized as follows. In Sec.~\ref{sec:particle_motion}, we present our formalism to compute GW waveforms for mergers in small mass-ratio regimes numerically. In Sec.~\ref{sec:qnm_model_fit} the fit of QNMs to the numerical waveforms is performed, which is necessary not only to extract the QNM parameters but also to infer the start time of ringdown after which the superposed-QNM modeling works well. Using the obtained parameters of the QNM model, in Sec.~\ref{sec:fisher_analysis} we conduct Fisher analysis to estimate the measurement errors for the mass and spin of an SMBH with LISA. Finally, we present the conclusion of our study in Sec.~\ref{sec:conclusion}. Note that for calculating the waveforms and fitting the QNMs, we use the geometrical unit, $G=c=1$. In physical units, the characteristic frequency $f_{\rm c}=c^3/2GM$ is related to the SMBH mass $M$ as
\begin{equation}
    f_{\rm c} \approx 10~{\rm mHz}\ \left(\frac{M}{10^7~M_\odot}\right)^{-1}.
\end{equation}

\section{GWs induced by a small compact object plunging into an SMBH }
\label{sec:particle_motion}
In this study, we consider GW signals induced by a small compact object falling into a spinning SMBH with mass $M$ and spin $a \coloneqq Mj$. We focus on the cases of $j=0.8, 0.9, 0.95$, and $0.99$ as it has been proposed that SMBHs of $M\sim 10^{7}M_\odot$ would have high spins in accordance with the observational findings (e.g.~\cite{Reynolds_2021, Inayoshi2024}) and cosmological simulations (e.g.~\cite{Sebastian_Springel_2019}). Denoting the mass of the object by $\mu$, we consider the case where the mass ratio, $q\coloneqq \mu/M$, is much smaller than unity.
Using Boyer-Lindquist coordinates, the background Kerr spacetime is given by
\begin{equation}
\begin{split}
    \diff s^2 =  & -\left(1-\frac{2Mr}{\Sigma}\right)\diff t^2 - \frac{4Mar\sin^2{\theta}}{\Sigma}\diff t \diff \phi + \frac{\Sigma}{\Delta} \diff r^2  \\ &+ \Sigma \diff \theta^2 + \left( r^2 + a^2 + \frac{2Ma^2 r}{\Sigma}\sin^2{\theta} \right)\sin^2{\theta}\diff \phi^2
\end{split} 
\end{equation}
where
\begin{align}
    &{\Delta(r) \coloneqq r^2 -2Mr + a^2} \\
    &{\Sigma(r) \coloneqq r^2 + a^2 \cos^2{\theta}.}
\end{align}
In this section, we describe how we obtain a GW waveform in the linear perturbation regime based on Ref.~\cite{Kojima-Nakamura}.

\subsection{Geodesic Trajectory of a Test Particle on the Equatorial Plane}
\begin{figure*}[t]
\centering
\includegraphics[scale=0.47]{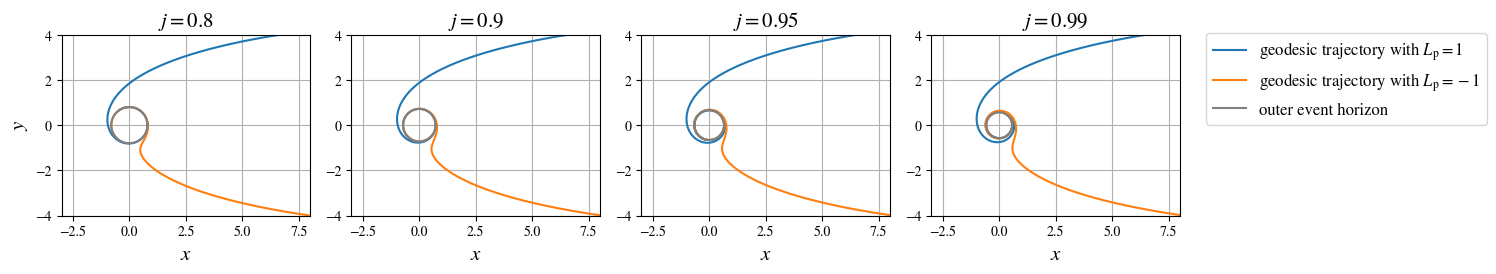}
\caption{Trajectories of a test particle with $L_\mathrm{p}=1$ (blue) and $L_\mathrm{p}=-1$ (orange) on the equatorial plane for the cases of $j=0.8, 0.9, 0.95$, and $0.99$ (from left to right). The SMBH is rotating counterclockwise in this figure, with the outer event horizon shown as a black circle. The coordinates are normalized such that $2M=1$.}
\label{fig:particle_trajectory}
\end{figure*}

By virtue of the separability of the Hamiltonian for particle's motion in the Kerr geometry~\cite{Carter_1968}, the trajectory of the particle plunging into a Kerr BH while restricted on the equatorial plane ($\theta = \pi/2$) is governed by the following equations:
\begin{gather}
\label{eq:eom_t}
    r^2 \frac{\diff t}{\diff \tau} = -a(a-L_\mathrm{p}) + \frac{r^2+a^2}{\Delta(r)}P(r)\:, \\
    \label{eq:eom_phi}
    r^2 \frac{\diff \phi}{\diff \tau} = -(a-L_\mathrm{p}) + \frac{a}{\Delta(r)}P(r)\:, \\
    \label{eq:eom_r}
    r^2 \frac{\diff r}{\diff \tau} = -\sqrt{Q(r)}\:,
\end{gather}
where
\begin{gather}
    P(r):= r^2 + a^2 - L_\mathrm{p} a\:, \\
    Q(r):= 2Mr^3 - {L_\mathrm{p}}^2 r^2 + 2Mr(L_\mathrm{p}-a)^2\:, 
\end{gather}
$\tau$ is the proper time of the particle and $L_\mathrm{p}$ is the orbital angular momentum of the particle normalized by its rest mass $\mu$. We here have taken the energy of the particle normalized by its rest mass $\mu$, $E_\mathrm{p} = 1$~\footnote{This corresponds to an approximation that the particle is at rest at infinity. We are interested in situations where a compact object, originally within the sphere of influence of the SMBH ($\lesssim$ pc for $M\sim 10^7~M_\odot$) but still far away, eventually plunges into the SMBH. The compact object sources ringdown GWs at the moment when it passes the light ring of the SMBH, which is also the case for the particle plunging into the hole from infinity with $E_{\rm p} =1$. Thus, our results shown below are expected to be qualitatively the same, as the trajectory at the light ring will nearly be insensitive to the initial velocity far away from the SMBH.}. and the Carter constant to be zero, which ensures that the particle is confined to the equatorial plane for the entire plunging orbit. In our calculations, we assume that the plunging object satisfies the infalling condition: $-2M(1+\sqrt{1+j})<L_\mathrm{p}<2M(1+\sqrt{1-j})$ and ignore the self-force of the object under the condition of $\mu \ll M$. Therefore, our waveforms presented here do not comprise any nonlinear effects, e.g., the back-reaction of the gravitational radiation to the light object's motion. In the following, we use the normalization of $2M=1$. 

Figure~\ref{fig:particle_trajectory} shows the plunging orbits of a test particle with $L_\mathrm{p}=1$  (blue) and $L_\mathrm{p}=-1$ (orange) for the cases of $j=0.8, 0.9, 0.95,$ and $0.99$ (from left to right). A particle with $L_\mathrm{p}=1$ travels counterclockwise, and the one with $L_\mathrm{p}=-1$ first travels clockwise, and afterward, it is inevitably dragged counterclockwise along the SMBH's rotation. 

\subsection{Calculation of the GWs Based on Sasaki-Nakamura Formalism}
\begin{figure*}[t]
\centering
\includegraphics[scale=0.47]{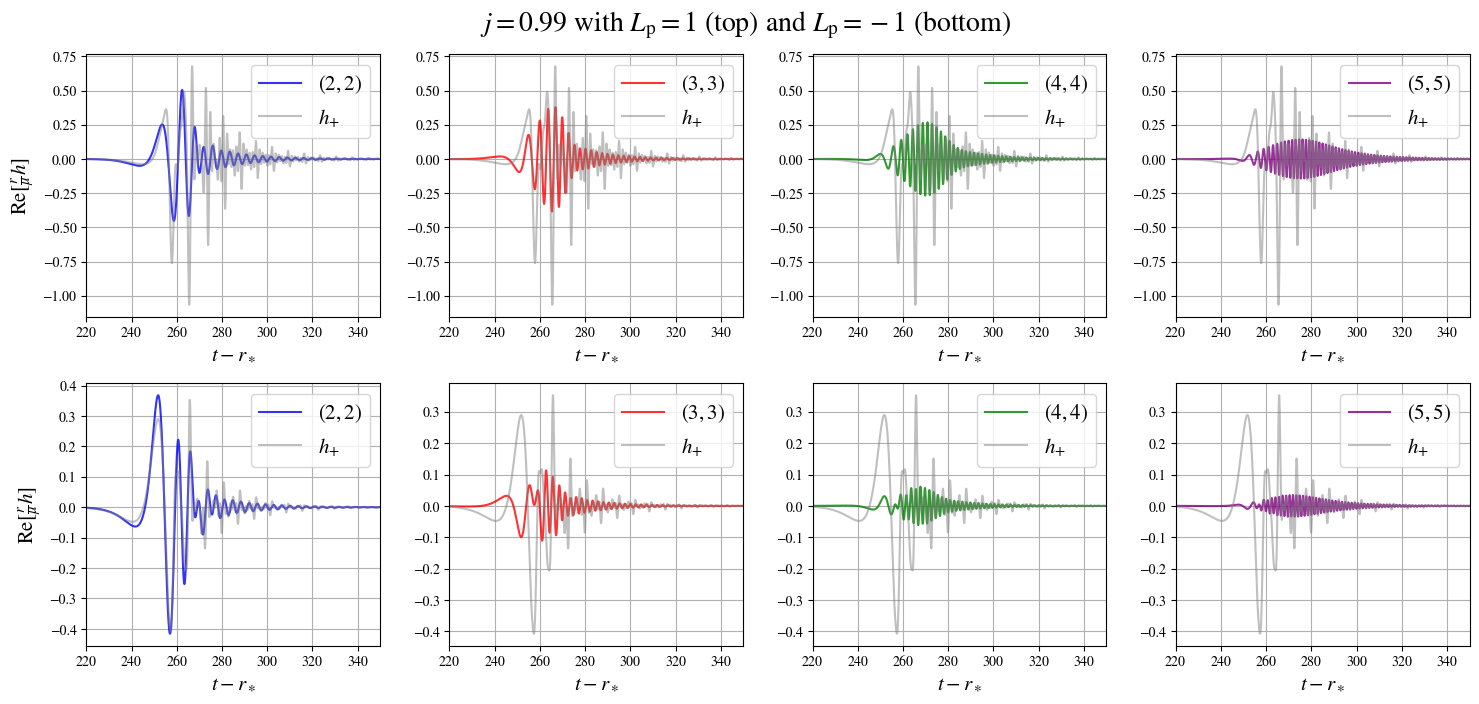}
\caption{Numerically calculated waveforms (normalized by $\mu$ and the distance $r$) of $(l,m)=(2,2), (3,3), (4,4),$ and $(5,5)$ modes (from left to right) for the case of $j=0.99$ with $L_\mathrm{p}=1$ (top) and $L_\mathrm{p}=-1$ (bottom). The horizontal axis indicates the retarded time. The grey line in each figure shows the total waveform up to the $(5,5)$ mode.}
\label{fig:gws_099}
\end{figure*}
The waveform induced by a small object traveling from infinity to the SMBH, as discussed in Sec.~\ref{sec:particle_motion}, can be numerically calculated based on the Sasaki-Nakamura (SN) formalism~\cite{Sasaki-Nakamura}. The perturbation variable of gravitational field, $X_{lm}$, follows the SN equation:
\begin{equation}
\label{eq:SN_eq}
    \left( \frac{\diff^2}{\diff {r_\ast}^2} - F_{lm}\frac{\diff}{\diff r_\ast} - U_{lm} \right) X_{lm}(\omega, r_\ast) = \tilde{T}_{lm}(\omega, r_\ast)\:,
\end{equation}
where $(l,m)$ denotes a label of the spheroidal harmonic mode, $r_\ast$ is the tortoise coordinate, $F_{lm}$ and $U_{lm}$ are functions of a radial coordinate whose explicit forms are given in Ref.~\cite{Sasaki-Nakamura}, and $\tilde{T}_{lm}$ is the source term associated with the geodesic motion of a plunging object whose expression is also given in Ref.~\cite{Kojima-Nakamura}. Adopting the Green function method, for a large $r_\ast$, Eq.~\eqref{eq:SN_eq} can be solved as
\begin{equation}
\begin{split}
    X_{lm}(\omega,r_\ast) & = X^\mathrm{(out)}_{lm}(\omega)\:e^{\mathrm{i}\omega r_\ast} \\ & = \int \tilde{T}_{lm}({r'_\ast}, \omega) G({r'_\ast}, r_\ast, \omega)\diff {r'_\ast}\:,
\end{split}
\end{equation}
where $G({r'_\ast}, r_\ast, \omega)$ is the Green's function constructed by the homogeneous solutions of the SN equation. The observed GW signal is expressed by a superposition of all multipole modes, 
\begin{gather}
    \tilde{h}(\omega) = \sum_{l,m}\tilde{h}_{lm}(\omega)\:, \\
    \tilde{h}_{lm}(\omega) := -\frac{2}{\omega^2}{}_{-2}S_{lm}(a\omega, \pi/2) R_{lm}(\omega)\:, \\
    R_{lm}(\omega) := \frac{-4 \omega^2 X^\mathrm{(out)}_{lm}}{\lambda_{lm}(\lambda_{lm}+2)-6\mathrm{i}\omega-12a^2\omega^2}\:,
\end{gather}
where ${}_{-2}S_{lm}$ is the spin-weighted spheroidal harmonics and $\lambda_{lm}$ is the eigenvalue determined by the regularity of ${}_{-2}S_{lm}$. The time domain waveform $h(t)$ can be obtained by the inverse Laplace transform of $\tilde{h}(\omega)$.

Figure~\ref{fig:gws_099} shows the numerically computed waveforms for $j=0.99$ and $L_{\rm p} = \pm 1$, decomposed to multipole modes from $(l,m)=(2,2)$ to $(5,5)$. Waveforms for other spin parameters are shown in Appendix.~\ref{ap:figures}. 
From those waveforms, one can see that the higher angular modes are excited more efficiently for rapidly spinning SMBHs, especially for $L_\mathrm{p}=1$.  As we show in Sec.~\ref{sec:result_fisher}, this results in higher angular modes having larger SNRs than the $(2,2)$ mode for rapid spins.

\section{QNM fitting}
\label{sec:qnm_model_fit}
For a given SMBH spin, the numerical waveform calculated above can be modeled as a sum of the QNMs of the SMBH at sufficiently late times. In this section we aim to extract the model parameters of the QNMs (their amplitudes and phases), that will be used as fiducial parameters for the Fisher analysis in Section \ref{sec:fisher_analysis}. 

In Sec.~\ref{sec:qnm_model_TD} we describe the methodology for the QNM fit, which aims to minimize the mismatch of the waveform and the QNM model for all the multiple modes. In Sec.~\ref{sec:result_mismatch}, we show the result of our QNM fits, for several spin parameters and two values of the particle's orbital angular momentum.

\begin{figure*}[t]
\centering
\includegraphics[scale=0.47]{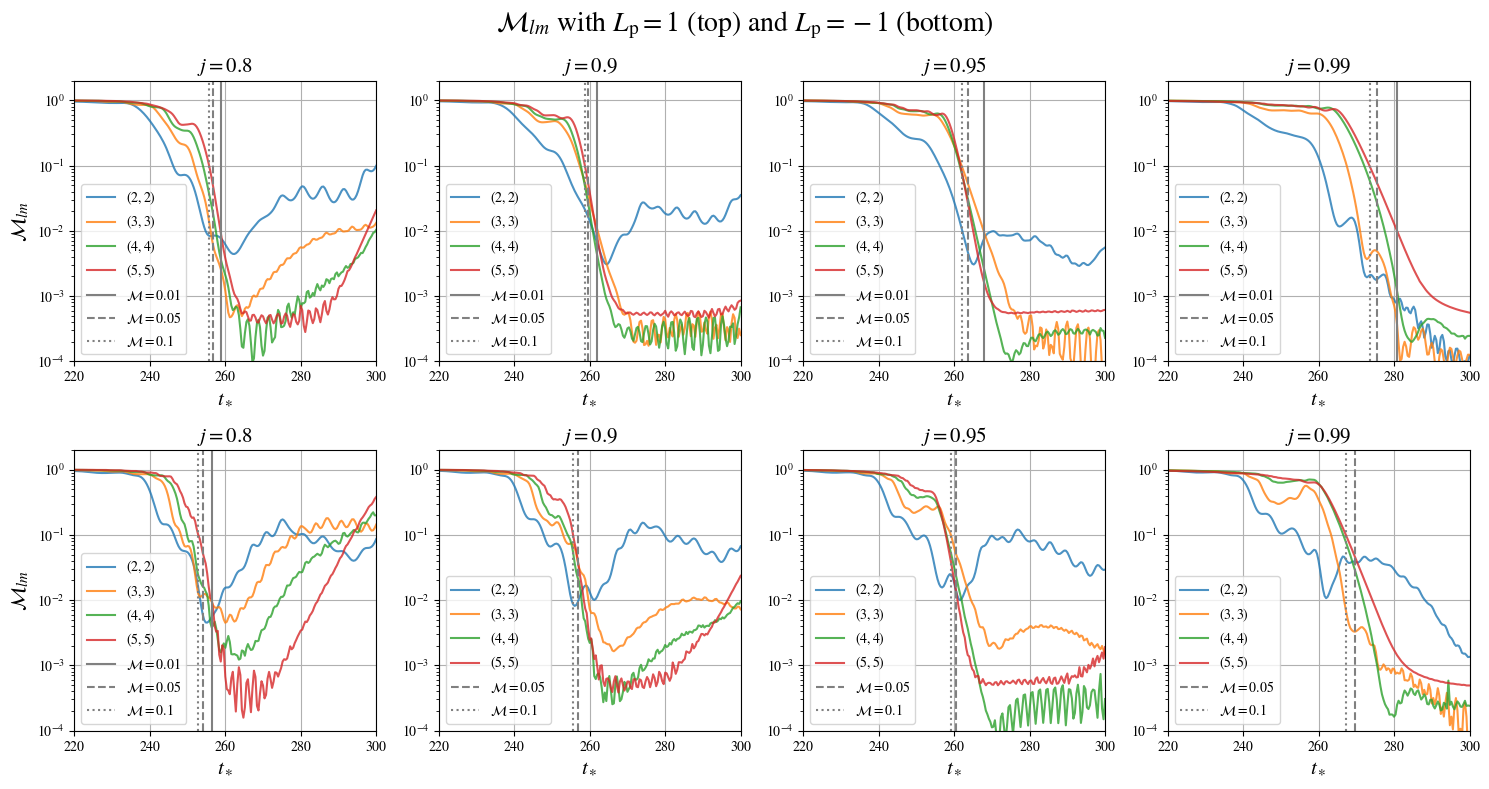}
\caption{Mismatch for the cases of $j=0.8, 0.9, 0.95,$ and $0.99$ with $L_\mathrm{p}=1$ (top) and $L_\mathrm{p}=-1$ (bottom). The grey vertical lines show the earliest time at which $\mathcal{M}_{lm}\leq 0.01$ (solid), $0.05$ (dashed), and $0.1$ (dotted), respectively, for all multipole modes. 
For the cases $j=0.9, 0.95,$ and $0.99$ with $L_\mathrm{p}=-1$ we only mark the times for $\mathcal{M}_{lm}<0.05$ and $0.1$, as $\mathcal{M}_{22}$ cannot be stably less than $0.01$ for a wide time range, or can only be less than $0.01$ at sufficiently late times. The former is because for several cases where the QNMs decay rapidly (e.g. $l=m=2$ modes), ${\cal M}_{lm}$ becomes larger at large values of $t_{\ast}$ as numerical noise dominates the data used in the computation of ${\cal M}_{lm}$.}
\label{fig:mismatch_all}
\end{figure*}

\begin{figure*}[t]
\centering
\includegraphics[scale=0.38]{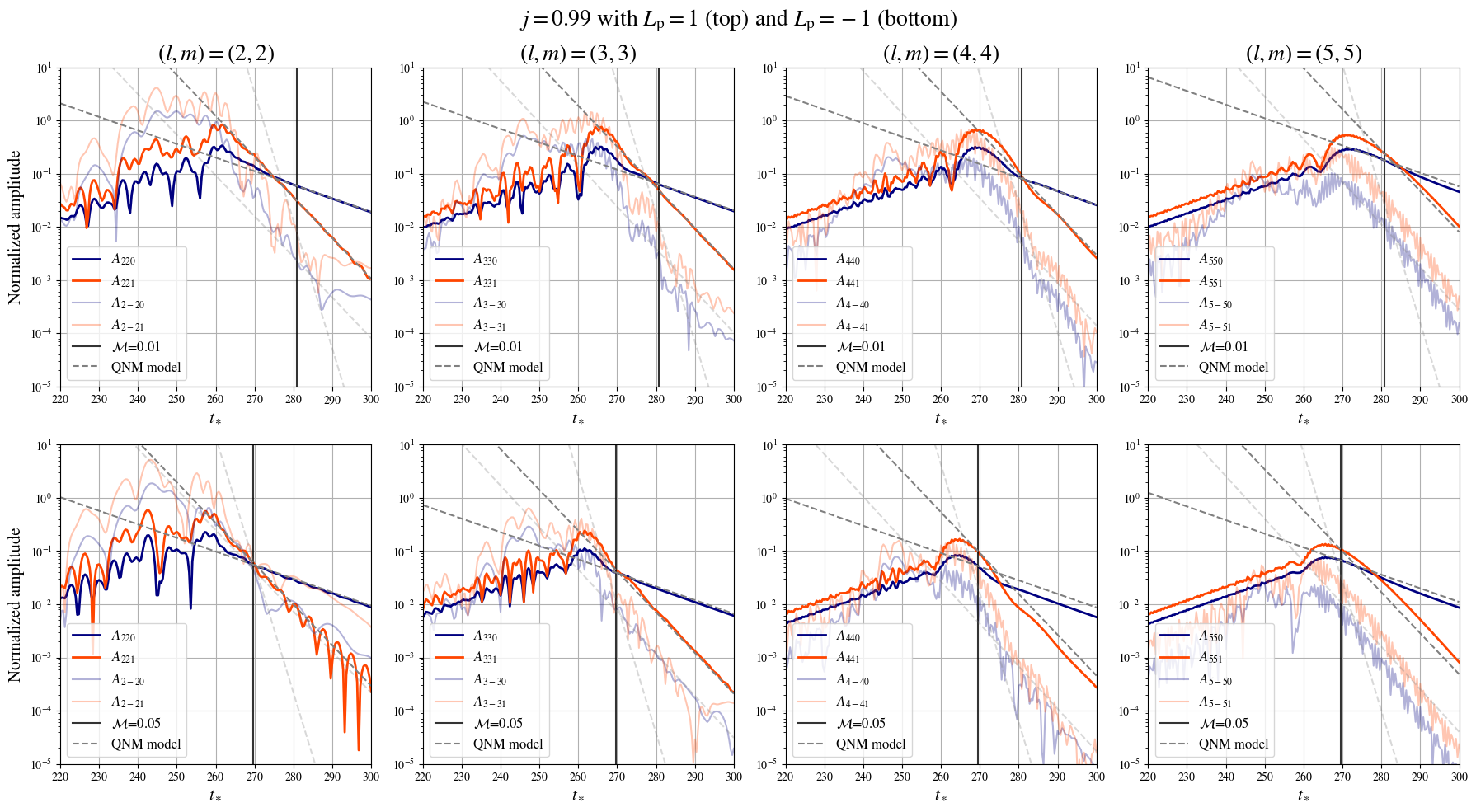}
\caption{Best-fit amplitudes with respect to $t=t_\ast$ for $j=0.99$ and for $l=m=2$, $3$, $4$ and $5$. We set $L_\mathrm{p}=1$ (top) and $L_\mathrm{p}=-1$ (bottom). We show the best-fit amplitudes $A_{lm0}$ (navy), $A_{lm1}$ (orange), $A_{l-m0}$ (light navy), and $A_{l-m1}$ (light orange). The grey dotted lines show the amplitudes of each QNM (Eq.~\eqref{eq:qnm_amp}) whose initial value is fixed at the earliest time at which $\mathcal{M}_{lm} \leq 0.01$ (upper panels) or $0.05$ (lower panels) for all multipole modes.} 
\label{fig:amp_099}
\end{figure*}

\subsection{Methodology}
\label{sec:qnm_model_TD}
For each multipole mode, we fit the QNM model to the numerical waveform in the time domain by using the least squares fit. The QNM model $h^{\mathrm{(QNM)}}_{lm}$ is given by
\begin{equation}
\label{eq:qnm_TD}
        h^{\mathrm{(QNM)}}_{lm}(t) = \sum_{n=0}^{n_\mathrm{max}} A_{lmn}\mathrm{e}^{-\mathrm{i}\omega_{lmn}(t-t_\ast) + \mathrm{i}\phi_{lmn}} \theta(t-t_\ast)\:,
\end{equation}
where $n_\mathrm{max}$ is the maximum overtone number in our model, $t_\ast$ is the start time of the ringdown, $\omega_{lmn}$ is the QNM frequency for $(l,m,n)$ mode, $A_{lmn}$ and $\phi_{lmn}$ are real-valued amplitude and phase of the mode at $t=t_\ast$, and $\theta(x)$ is the step function. Here, $\{A_{lmn},\phi_{lmn}, t_*\}$ are the fitting parameters for our numerically calculated waveform. Throughout this study, we use \texttt{qnm}, a python module to calculate the QNM frequencies $\omega_{lmn}$~\cite{Stein_2019}. 

For each multipole mode of $(l,m)$, we adopt 
\begin{equation}
    \{(l,m,n)\} = \{(l,m,0), (l,m,1), (l,-m,0), (l,-m,1)\}.
\end{equation}
To avoid overfitting due to the inclusion of many QNMs in the model, we consider $n_{\rm max}=1$ in this study. The $(l,-m,n)$ modes are called mirror modes, and the excitations of these modes are known to be suppressed for mergers of BHs with comparable masses ~\cite{mirror_modes_2020, Dhani_Sathya_2021}. We nevertheless took them into account, as it is less clear whether they are excited for mass ratios of our interest.

For each multiple mode, we assess the goodness-of-fit by evaluating the mismatch ${\cal M}_{lm}$ defined by
\begin{equation}
    \mathcal{M}_{lm}(t_\ast) := 1 -\frac{|(h_{lm}(t)|h^\mathrm{(QNM)}_{lm}(t))|}{\sqrt{(h_{lm}(t)|h_{lm}(t))(h^\mathrm{(QNM)}_{lm}(t)|h^\mathrm{(QNM)}_{lm}(t))}}\:.
\end{equation}
The inner product $(A(t)|B(t))$ is defined by
\begin{equation}
    (A(t)|B(t)) := \int^{\infty}_{t_\ast} A^\ast(t) B(t) \diff t,
\end{equation}
which depends on the assumed start time of ringdown $t_{\ast}$. The superscript ${}^{\ast}$ indicates the complex conjugate. For a given $t_{\ast}$, we obtain the best-fit QNM parameters $\{A_{\rm lmn}, \phi_{lmn}\}$ as well as the mismatch $\mathcal{M}$. If the amplitude $A_{lmn}$ is extracted self-consistently, $A_{lmn}(t_*)$ should follow an exponential decay
\begin{equation}
\label{eq:qnm_amp}
    A_{lmn}(t_\ast) \propto \mathrm{exp}\left[-\frac{t_\ast}{\tau_{lmn}}\right]\:,
\end{equation}
where $\tau_{lmn}:= \{-\mathrm{Im}(\omega_{lmn})\}^{-1}$ is the damping time for the $(l,m,n)$ mode. Comparing $A_{lmn}(t_*)$ to this relation enables a cross-check that our QNM fitting is valid at sufficiently late times when the relevant QNMs in the fit dominate the ringdown signal.

\subsection{Results}
\label{sec:result_mismatch}
We here perform the QNM fit for the same cases as in Sec.~\ref{sec:particle_motion}. We also consider the case of $L_\mathrm{p}=\pm 1$ for each spin $j$ to check the impact of the choice of $L_\mathrm{p}$ on the error estimates.

Figure~\ref{fig:mismatch_all} shows the values of mismatch with respect to the assumed start time of ringdown $t_*$. One can see that the mismatch decays earlier for lower multipole modes. It is consistent with the proposals~\cite{Oshita_2023, Oshita_Tsuna_2023} that more overtones are excited for higher multipole modes based on the computation of the excitation factors. In all cases, the higher angular modes have better fits than the $(2,2)$ mode due to their larger quality factors, i.e., larger values of $|\text{Re}(\omega_{lmn})/\text{Im}(\omega_{lmn})|$ that enable more robust extraction of individual QNM from the numerical waveforms.

Figure~\ref{fig:amp_099} shows the best-fit amplitudes with respect to $t=t_\ast$ for the case of $j=0.99$. The figures for other spin cases are shown in Appendix.~\ref{ap:figures}. 
In all cases, the extracted amplitude for the fundamental modes ($n=0$) follows Eq.~\eqref{eq:qnm_amp}, which implies that the amplitude is extracted in a self-consistent manner. For the higher spin cases, the first overtone of the higher multipole modes also follows Eq.~\eqref{eq:qnm_amp}. These fitting results indicate that a large quality factor helps to stably extract the QNM amplitudes. On the other hand, the stable extractions of the mirror modes are not achieved for all cases with $L_\mathrm{p}=1$ considered here. However, for the cases with $L_\mathrm{p}=-1$ the fundamental mirror modes for higher multipole modes are barely extracted at late times (se Fig.~\ref{fig:amp_099}). In addition, as shown in Figs.~\ref{fig:amp_08}, \ref{fig:amp_09}, and \ref{fig:amp_095} in Appendix~\ref{ap:figures}, the mirror modes are barely extracted for several cases with negative $L_\mathrm{p}$, such as the $(4,4)$ and $(5,5)$ modes in the case of $j=0.8$. These facts indicate that infalling objects with negative $L_\mathrm{p}$ probably excite the mirror modes to some extent.

\section{Measurement Errors of mass and spin of an SMBH}
\label{sec:fisher_analysis}
With future GW detections of extreme mass ratio ($q\ll 1$) merger events by LISA in mind, we estimate the expected statistical errors of the (source-frame) SMBH mass $M$ and spin $j$ for such mergers, based on the Fisher matrix formalism~\cite{Cutler_Flanagan_1994, Poisson_Will_1995, Flanagan_Hughes_1998, Berti_Cardoso_Will_2006} appropriate for events with high SNR. 

Ref.~\cite{Berti_Cardoso_Will_2006} has discussed the errors with analytical forms under several assumptions], most importantly the GW amplitude emitted by each QNM being an uncertain free parameter. Keeping some of the assumptions made in Ref.~\cite{Berti_Cardoso_Will_2006}, we estimate the errors in the mass-spin inference from GWs sourced by an extreme mass ratio merger, where we use the amplitudes obtained from first principles using our numerical waveform and their QNM fits.

In Sec.~\ref{sec:fisher_matrix_formalism}, a brief review of the Fisher matrix formalism is given. We then apply the Fisher analysis to our QNM model in Sec.~\ref{sec:qnm_mode_fisher}, which consists only of $(l,m,0)$ and $(l,m,1)$ modes up to $(l,m)=(5,5)$ mode. The fiducial values of $A_{lmn}, \phi_{lmn}$, and $t_\ast$ are fixed based on the fitting results shown in the previous section. To calculate the errors with the multiple modes, we will employ a simplified form of the Fisher matrix which can be assumed under an approximate relation among the spin-weighted spheroidal harmonics~\cite{Berti_Cardoso_Casals_2006}. 
In Sec.~\ref{sec:result_fisher}, we show the results of our Fisher analysis.
The dependence of the root-mean-square (RMS) errors on $j$, $L_\mathrm{p}$, $q$, $d_\mathrm{L}$ and $M$ is shown, where $d_\mathrm{L}$ is the luminosity distance calculated based on the cosmological parameters provided in Ref.~\cite{Planck_2020}. Also, we demonstrate how efficiently the combination of the several multipole modes reduces the statistical errors. 

\subsection{Fisher Matrix Formalism}
\label{sec:fisher_matrix_formalism}
The Fisher matrix $\Gamma$ is given by 
\begin{equation}
    (\Gamma)_{ij} := \langle \partial_i \tilde{h}(f;\boldsymbol{\theta}) | \partial_j \tilde{h}(f;\boldsymbol{\theta}) \rangle\:,
\end{equation}
where $\{\boldsymbol{\theta}\}$ is a parameter set in a GW model, $\tilde{h}(f;\boldsymbol{\theta})$ is a GW model in frequency domain, and $\partial_i$ is the partial derivative with respect to parameter $\theta_i$. The inner product $\langle A(f) | B(f) \rangle$ is defined by
\begin{equation}
\label{eq:inner_product_FD}
    \langle A(f) | B(f) \rangle := 4\:\mathrm{Re} \left[ \int^{f_\mathrm{max}}_{f_\mathrm{min}} \frac{A^\ast(f) B(f)}{S(f)} \diff f \right]\:,
\end{equation}
where $A(f)$ and $B(f)$ are functions of $f$. $S(f)$ is the full noise curve for LISA~\cite{LISA_noise_2019} averaged over the sky position and polarization (the concrete form shown in Appendix.~\ref{ap:LISA_noise}). We set $f_\mathrm{min} = 10^{-4}$ Hz which is a conservative choice for the LISA operation, and we take $f_\mathrm{max} = 1$ Hz, which is large enough for the convergence of the integration in the frequency domain. The RMS errors of $\theta_i$, denoted by $\delta \theta_i$, and the correlation coefficient between $\theta_i$ and $\theta_j$, denoted by $c_{\theta_i \theta_j}$, are estimated by
\begin{gather}
    \delta \theta_i = \sqrt{(\Gamma^{-1})_{ii}}\:,\\
    c_{\theta_i \theta_j} = \frac{(\Gamma^{-1})_{ij}}{\sqrt{(\Gamma^{-1})_{ii} (\Gamma^{-1})_{jj}}}\:,
\end{gather}
where $\Gamma^{-1}$ is the inverse matrix of $\Gamma$.

\subsection{Error Estimations with Multiple Angular Modes}
\label{sec:qnm_mode_fisher}
We do not include the mirror modes here because they are subdominant and not extracted stably for all cases as shown in Sec.~\ref{sec:result_mismatch}. Thus, our QNM model for $(l,m)$ mode in the time domain is given by
\begin{equation}
\label{eq:qnm_TD_fisher}
\begin{split}
    h^{\mathrm{(QNM)}}_{lm}(t) = \:&\big( A_{lm0}\mathrm{e}^{ -\mathrm{i}\omega_{lm0}(t-t_\ast) +\mathrm{i}\phi_{lm0} } \\ &+ A_{lm1}\mathrm{e}^{ -\mathrm{i}\omega_{lm1}(t-t_\ast)  +\mathrm{i}\phi_{lm1} }\big) \theta(t-t_\ast)\:.
\end{split}
\end{equation}
The QNM model for $(l,m)$ mode in the frequency domain is derived by the Fourier transformation,
\begin{equation}
\label{eq:qnm_FD}
    \tilde{h}^{\mathrm{(QNM)}}_{lm}(\omega) =\frac{\mathrm{i}}{2 \pi} \left( \frac{A_{lm0} \mathrm{e}^{\mathrm{i}\omega t_\ast + \mathrm{i}\phi_{lm0}}}{\omega-\omega_{lm0}} + \frac{A_{lm1} \mathrm{e}^{\mathrm{i}\omega t_\ast + \mathrm{i}\phi_{lm1}}}{\omega-\omega_{lm1}}  \right)\:.
\end{equation}
Therefore, our QNM model is given by
\begin{equation}
    \tilde{h}^{\mathrm{(QNM)}}(\omega) = \sum_{lm} \tilde{h}^{\mathrm{(QNM)}}_{lm}(\omega)\:,
\end{equation}
which has 18 model parameters in total, denoted by $\{ \boldsymbol{\theta} \}$,
\begin{equation}
    \{\boldsymbol{\theta} \} = \{M,\:j,\:\underset{lmn}{\cup}\{ \mathcal{A}_{lmn},\:\phi_{lmn}\}\}\:,
\end{equation}
where $\mathcal{A}_{lmn}$ is defined by
\begin{equation}
\label{eq:A_lmn}
    \mathcal{A}_{lmn} := \frac{q M_z}{d_\mathrm{L}} A_{lmn}\:,
\end{equation}
and $M_z:=(1+z)M$ is the redshifted BH mass.

The value of $t_\ast$ in Eq.~\eqref{eq:qnm_FD} is chosen at the earliest time when $\mathcal{M}_{lm} \leq 0.05$ ($L_\mathrm{p}=-1$) or $\mathcal{M}_{lm} \leq 0.01$ ($L_\mathrm{p}=1$) is satisfied for all multipole modes. The fiducial values for $A_{lmn}$ and $\phi_{lmn}$ are set to the best-fit values at $t=t_\ast$ for each mode.
However, we have confirmed that the final RMS errors for $M$ and $j$ are nearly insensitive to the choice of the upper bound on ${\cal M}_{lm}$ as long as it is $\lesssim 0.1$, whose details are discussed in Appendix.~\ref{ap:insensitivity}. Note that in the full Bayesian data analysis, the misalignment between the QNM model and the real signal can lead to systematic bias and leads to the deviation between the estimated values and \textit{true} values. 
We here assume that the imposed upper bounds on ${\cal M}_{lm}$ are small enough not to cause the bias in the data analysis. 

As we have seen in the previous sections, significant excitations of higher multipole modes are expected for an extreme mass-ratio merger involving a rapidly spinning SMBH. We thus calculate the expected errors of $M$ and $j$ by taking into account higher multipole modes as well.
As discussed in Sec.~VI in Ref.~\cite{Berti_Cardoso_Will_2006}, considering the angular-averaged case, the Fisher matrix can be simplified as\footnote{When calculating $\Gamma$, we convert the QNM model parameters from the geometrical units to physical units by putting back the dimensional constants $G$, $c$ and $2M$.}
\begin{equation}
    \Gamma = \Gamma_{22} + \Gamma_{33} + \Gamma_{44} + \Gamma_{55}\:,
\end{equation}
where $\Gamma_{lm}$ is the Fisher matrix for each $(l,m)$ mode. This approximation is justified by the approximated relation for the scalar product of spheroidal harmonics~\cite{Berti_Cardoso_Casals_2006},
\begin{equation}
    \int S^\ast_{lmn}(\theta, \phi) S_{l'm'n'}(\theta, \phi) \diff \Omega \simeq \delta_{ll'} \delta_{mm'}\:,
\end{equation}
where $\diff \Omega$ is the infinitesimal solid angle, and $\delta_{ij}$ is the Kronecker's delta.

\subsection{Results}

\label{sec:result_fisher}
\begin{table*}[t]
\begin{ruledtabular}
\caption{\label{tb:error_1}%
SNRs, RMS errors for $M$ and $j$, and the correlation coefficient between $M$ and $j$ for the case of $L_\mathrm{p}=1$. The start time of ringdown $t_\ast$ and the amplitude $A_{lmn}$ are fixed at the earliest time at which $\mathcal{M}_{lm} \leq 0.01$ for all modes. Here we fix $M=10^{7}M_\odot, q=10^{-3}$, and $d_\mathrm{L}=1\:\mathrm{Gpc}$.  
}
\begin{tabular}{cccccccc}
 BH spin & SNR of $(2,2)$ & SNR of $(3,3)$ & SNR of $(4,4)$ & SNR of $(5,5)$ & $\delta M/M [\%]$ & $\delta j/j [\%]$ & $c_{\mathrm{log}{M}\mathrm{log}j}$ \\
\colrule
$j=0.99$ & 19.8 & 22.9 & 32.6 & 47.7 & $0.328$ & $0.0600$ & $-0.985$ \\
$j=0.95$ & 22.3 &
59.7 & 42.4 & 16.3
 & $0.433$ & $0.201$ & $-0.967$ \\
$j=0.9$ & 73.0 &
63.4 & 26.9 & 12.4 & $0.563$ & $0.406$ & $-0.969$ \\
$j=0.8$ & 63.7 &
34.0 & 
16.3 & 8.13 & $1.29$ & 1.63 & $-0.969$
\end{tabular}
\end{ruledtabular}
\end{table*}
\begin{table*}[t]
\begin{ruledtabular}
\caption{\label{tb:error_m1}%
Same as Table~\ref{tb:error_1}, but $L_\mathrm{p}=-1$.}
\begin{tabular}{cccccccc}
 BH spin & SNR of $(2,2)$ & SNR of $(3,3)$ & SNR of $(4,4)$ & SNR of $(5,5)$ & $\delta M/M [\%]$ & $\delta j/j [\%]$ & $c_{\mathrm{log}{M}\mathrm{log}j}$ \\
\colrule
$j=0.99$ & 19.7 & 13.5 & 16.2 & 14.8 & $0.647$ & $0.112$ & $-0.978$ \\
$j=0.95$ & 51.8 &
24.5 & 11.9 & 4.66
 & $0.919$ & $0.394$ & $-0.965$ \\
$j=0.9$ & 71.6 &
19.7 & 6.07 & 2.89 & $1.14$ & $0.734$ & $-0.952$ \\
$j=0.8$ & 57.0 &
10.8 & 
3.46 & 1.58 & $2.51$ & 2.83 & $-0.952$
\end{tabular}
\end{ruledtabular}
\end{table*}

\subsubsection{\textbf{Dependence of $j$ and $L_\mathrm{p}$ on the Errors}}
\label{sec:result_fisher_j_L}
We estimate the SNRs, RMS errors for $M$ and $j$, and the correlation coefficient between $M$ and $j$. We here fix luminosity distance $d_\mathrm{L}=1\:\mathrm{Gpc}$ and the mass of an SMBH at the source frame $M=10^7M_\odot$, which are the same choices as in Ref.~\cite{Oshita_Tsuna_2023}.  Table~\ref{tb:error_1} and \ref{tb:error_m1} show the results for $L_\mathrm{p}=1$ and $L_\mathrm{p}=-1$, respectively.

We find that the more rapidly the SMBH rotates, the smaller the errors for $M$ and $j$ are. This is qualitatively consistent with the analytic error estimation in Ref.~\cite{Berti_Cardoso_Will_2006}. In particular, $j$ is expected to be more precisely estimated than $M$ ($\delta j /j \lesssim 1\:\mathrm{\%}$) for the cases of $j>0.9$. Also, the expected errors of the remnant parameters for the case of $L_\mathrm{p}=1$ and $L_\mathrm{p}=-1$ differ by factor $\lesssim 2$, though the SNRs can differ by several times\footnote{The SNR of the $(2,2)$ mode for the case of $j=0.95$ is larger than that for $L_\mathrm{p}=1$, which is due to the unstable fit of the amplitude shown in Fig.~\ref{fig:amp_095}. We find that this does not affect the final result, since excluding the $(2,2)$ mode in the fit changes the measurement precisions by only $\lesssim 20$\%.}.

This implies that the errors for the remnant parameters are not so sensitive to the orbital angular momentum $L_\mathrm{p}$ of the plunging particle. Furthermore, the tables show that the correlation coefficient between $M$ and $j$ takes a similar value $-0.98 \sim -0.95$ (negative correlation) for all cases, which can be interpreted as a model bias. The sign is different from the results given in Ref.~\cite{Berti_Cardoso_Will_2006}, which is due to the different definitions of the QNM model in the frequency domain.

\subsubsection{\textbf{Dependence of $q$ on the Errors}}
The amplitude of the signal and the SNR both scale with the mass ratio of the binary $q$ (see Eq.~\eqref{eq:A_lmn}). While the SNR values scale with $q$ by definition, the errors and the correlation coefficients generally do not have such scaling. Nevertheless, we find that for the case of $q=10^{-5}$ (other quantities are set to the same values as Table~\ref{tb:error_1}), the measurement errors simply become 100 times the results shown in Table~\ref{tb:error_1}, similar to the SNR values. This is probably because the LISA noise curve is nearly constant around the QNM frequencies of BHs with masses of $\sim 10^{7}M_\odot$.

\subsubsection{\textbf{Comparison with Single Angular Mode Analysis}}
\label{sec:result_fisher_comparison}
We performed the Fisher matrix analysis by combining several multipole modes and showed the results in Sec.~\ref{sec:result_fisher_j_L}. To check the efficiency of combining several multipole modes, we here conduct a similar Fisher analysis of the RMS errors when measuring an individual multipole mode. 
The QNM model for each multipole mode Eq.~\eqref{eq:qnm_FD} now contains only 6 model parameters:
\begin{equation}
    \{\boldsymbol{\theta}_{lm} \} = \{M,\:j,\: \:\underset{n}{\cup}\{\mathcal{A}_{lmn},\:\phi_{lmn}\}\}\:.
\end{equation}
With this model, we calculate the inverse matrix of $\Gamma_{lm}$ and estimate the errors with a single multipole mode. Table.~\ref{tb:error_22} shows the errors and the correlation coefficients for $j=0.8, 0.9, 0.95$ and $0.99$ with $L_\mathrm{p}=1$. Comparison with the values in Table.~\ref{tb:error_1} indicates that the errors of $M$ and $j$ in the multimode analysis are mostly determined by $(3,3)$ modes (and also by $(5,5)$ mode for $j=0.99$), while improvements in precision of up to a factor of a few can be seen by the combination with other angular modes. 

These results imply that a precise measurement of the remnant parameters can be achieved when many angular modes are significantly excited. Furthermore, the larger quality factors (due to higher QNM frequencies) in higher multipole modes considerably contribute to the reduction of the errors. This is analogous to the parameter estimation in the inspiral phase: the number of orbital cycles in band can play a crucial role in precisely determining the chirp mass of a binary~\cite{Cutler:1994ys, Poisson:1995ef}.  

\begin{table}[h]
\begin{ruledtabular}
\caption{\label{tb:error_22}%
RMS errors for $M$ and $j$, and the correlation coefficient between $M$ and $j$ estimated by a single angular mode, for $j=0.8, 0.9, 0.95$ and $0.99$ with $L_\mathrm{p}=1$.}
\begin{tabular}{ccccc}
BH spin & $(l,m)$ &  $\delta M/M [\%]$ & $\delta j/j [\%]$ & $c_{\mathrm{log}{M}\mathrm{log}j}$ \\
\colrule
$j=0.8$ & $(2,2)$ &  3.48 & 3.55 &
 $-0.956$  \\
${}$ & $(3,3)$ & 1.56 & 2.06 &
 $-0.972$ \\
${}$ & $(4,4)$ & 6.60  & 7.96 &
 $-0.986$ \\
${}$ & $(5,5)$ & 11.1 & 13.5 &
 $-0.990$ \\
 \colrule
 
$j=0.9$ & $(2,2)$ &  2.78 & 1.53 &
 $-0.952$  \\
${}$ & $(3,3)$ & 0.608 & 0.446 &
 $-0.969$ \\
${}$ & $(4,4)$ & 2.76 & 1.92 &
 $-0.986$ \\
${}$ & $(5,5)$ & 5.37 & 3.79 &
 $-0.990$ \\
\colrule

$j=0.95$ & $(2,2)$ & 6.13 & 2.41 &
 $-0.974$  \\
${}$ & $(3,3)$ & 0.465 & 0.222 &
 $-0.959$ \\
${}$ & $(4,4)$ & 1.33 & 0.585 &
 $-0.987$ \\
${}$ & $(5,5)$ & 3.90 & 1.74 &
 $-0.992$ \\
 \colrule
 
$j=0.99$ & $(2,2)$ &  2.87 & 0.414 &
 $-0.971$  \\
${}$ & $(3,3)$ & 0.615 & 0.129 &
 $-0.988$ \\
${}$ & $(4,4)$ &  1.64 & 0.272 &
 $-0.990$ \\
${}$ & $(5,5)$ & 0.529 & 0.0914 &
 $-0.990$
\end{tabular}
\end{ruledtabular}
\end{table}

\subsubsection{\textbf{Dependence of $d_\mathrm{L}$ and $M$ on the Errors}}
\label{sec:result_fisher_dL_M}
\begin{figure*}[t]
\centering
\includegraphics[scale=0.47]{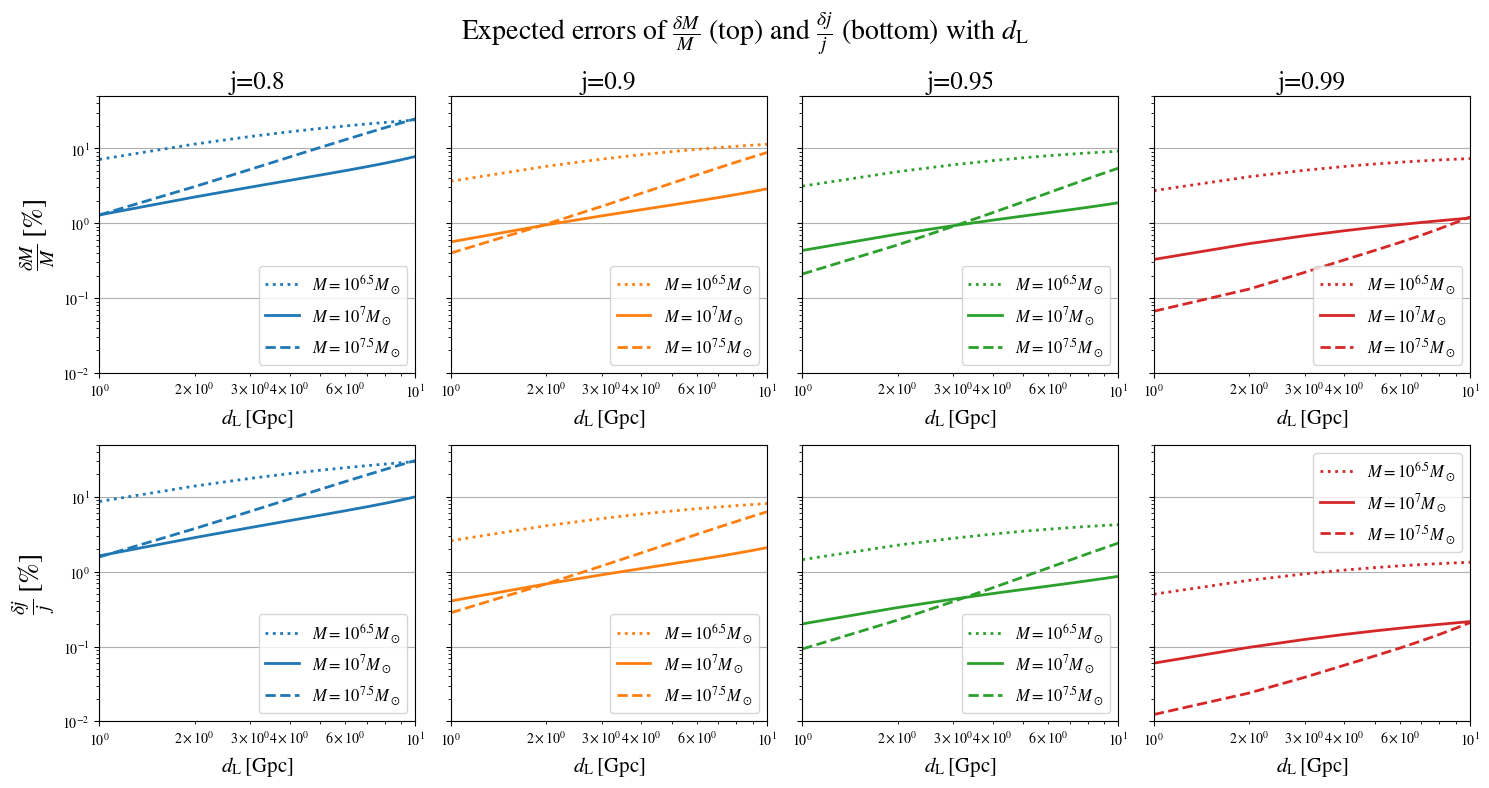}
\caption{RMS errors of $\delta M/M$ (top) and $\delta j/j$ (bottom). Each panel shows the cases of the source frame BH mass $M=10^{6.5}M_\odot$ (dotted), $M=10^{7}M_\odot$ (solid), and $M=10^{7.5}M_\odot$ (dashed). The mass ratio $q$ is fixed to $10^{-3}$.}
\label{fig:errors_Mj_dL}
\end{figure*}

Finally, we investigate the dependence of the RMS errors for the remnant parameters on $d_\mathrm{L}$ and $M$. Figure~\ref{fig:errors_Mj_dL} shows the expected errors of $\delta M/M$ (top) and $\delta j/j$ (bottom). We estimate the errors for up to $d_\mathrm{L} = 10\:\mathrm{Gpc}$ (corresponding to $z=1.37$), and consider $M=\{10^{6.5},10^7,10^{7.5}\}M_\odot$ within the mass range that we expect to typically have rapidly spinning SMBHs \citep{Sebastian_Springel_2019}.

We find that for intermediate (rapid) spins, the remnant parameters of an SMBH with $\sim 10^{7} M_{\odot}$ ($\sim 10^{7.5} M_{\odot}$) can be most precisely measured.
Depending on the luminosity distance $d_\mathrm{L}$, mass $M$ and spin $j$ of a SMBH, the typical GW frequency of the ringdown changes due to cosmological redshift. This causes such a non-monotonic distance-dependent dependence of the measurement errors on SMBH mass.

For the range of $d_\mathrm
{L}$ we cover, the errors for $M=10^{6.5}M_\odot$ are generally larger than that for $10^{7}$--$10^{7.5}  M_{\odot}$. This is a combination of an intrinsically weaker signal, and the frequencies of QNMs for $M=10^{6.5}M_\odot$ being higher that it tends to be more buried in the LISA noise when $z$ is small. However, for more distant sources the QNMs for $M=10^{6.5}M_\odot$ can enter the most sensitive frequency band due to cosmological redshift. This results in precision comparable to the $10^{7.5}M_\odot$ case for $d_{\rm L}\approx 10$ Gpc.

\subsection{Discussion}
\label{sec:discussion}
In this section we discuss the limitations and future avenues of our work, that may be improved by future studies.

An important assumption in our model is neglecting the self-force of the lighter object when calculating the GW waveform. Given that the self-force affects the signal (as well as the final mass and spin of the SMBH) on the order of $q$, we cannot reliably predict precisions better than $\mathcal{O}(q)$ with our waveform model. Therefore, our results are reliable if $\delta \theta_i \gtrsim q$ is satisfied, but for $\delta\theta_i \lesssim q$ we may need to consider systematic uncertainties due to the waveform modeling. In the cases we investigated for $q=10^{-3}$, this only matters for the near-extremal SMBHs at $d_{\rm L}\lesssim$ a few Gpc (see Fig.~\ref{fig:errors_Mj_dL}), and is even less important for smaller mass ratios. Treatment to include these effects is left for future study.

The case of $M\sim 10^{7}M\odot$ and $q\sim10^{-3}$ or $10^{-5}$ corresponds to a merger between an SMBH and an intermediate-mass BH (IMBH). The event rate of SMBH--IMBH mergers is uncertain, but recent N-body simulations suggest a range $0.003$-$0.03\:\mathrm{Gpc}^{-3}\mathrm{yr}^{-1}$~\cite{Fragione:2022egh, Panamarev:2018bwq} or $2$-$20~\mathrm{yr}^{-1}$ within $z<1$ ($d_\mathrm{L}\lesssim 7\:\mathrm{Gpc}$)~\cite{Amaro-Seoane_2023}. The mergers might be related to the formations of EMRI/IMRIs, where the secondary BH undergoes many inspirals with very high eccentricities before plunging into the BH. Detailed estimates of the event rate as a function of SMBH/IMBH masses are beyond the scope of this work, but combining such rate predictions with our work would enable a more realistic prospect of constraining the properties of SMBHs with LISA.

There is another way to parametrize the QNM model. It utilizes the quality factor, $Q_{lmn}:=\pi f_{lmn} \tau_{lmn}$, rather than $M$ and $j$, and the model is given by
\begin{equation}
    h^{\mathrm{QNM}}_{lm}(t) = \sum_{lmn} A_{lmn} e^{-\frac{\pi f_{lmn}}{Q_{lmn}}t} e^{-2i\pi f_{lmn} t + i\phi_{lmn}}\:.
\end{equation}
The model parameters here are $\{ \boldsymbol{{\theta}} \}= \underset{lmn}{\cup} \{ f_{lmn}, Q_{lmn}, A_{lmn}, \phi_{lmn} \}$.
Using this model we can infer the mass and spin by estimating the multiple peaks of $Q_{lmn}$ in the posteriors and identifying possible combinations of the mass and spin (see Sec.~VI.C in Ref.~\cite{Berti_Cardoso_Will_2006} for details). While we have focused on the error estimates introduced in Sec.~\ref{sec:fisher_matrix_formalism} following a more straightforward Fisher matrix analysis, it might be important as future work to consider the prospects of this approach as a complementary way to constrain the SMBH properties.

\section{Conclusion}
\label{sec:conclusion}
In anticipation of future detections of ringdown gravitational waves (GWs) from supermassive black holes (SMBHs) by LISA, we obtained the estimates of the statistical errors for the mass and spin of SMBHs by such detections, focusing on SMBH masses of $\sim 10^{7}M_\odot$ where LISA is most sensitive. We have numerically computed the ringdown GWs induced by a low-mass compact object plunging into an SMBH, and extracted the amplitudes for the fundamental mode and the first overtone. We then used the Fisher matrix formalism to estimate the measurement errors of the SMBH's mass and spin by observing such events with LISA, for a broad range of SMBH mass/spin and luminosity distance.

The main result is that a more precise estimation of the mass and spin is expected for more rapid SMBH spins. In particular, for a merger involving a near-extremal ($j=0.99$) SMBH of mass $\sim 10^7M_\odot$ with mass ratio of $10^{-3}$ at $\lesssim 10\:\mathrm{Gpc}$ (corresponding to $z \lesssim 1.37$), the measurement error in the mass and spin would be $\sim 1\:\mathrm{\%}$ and $\sim 10^{-1}\:\mathrm{\%}$, respectively. We can expect precise spin measurements at the percent level for SMBHs of $j\gtrsim 0.9$ at these distances, which will enable us to confirm (or refute) their high-spinning nature.

As discussed in Sec.~\ref{sec:discussion}, we note that we cannot reliably predict measurements more precise than of the order of $q$ ($\sim 0.1~\%$ for $q=10^{-3}$), since self-force effects are not included in our calculation. Obtaining precision beyond this limit requires taking into account such effects, which is left for a future study. For an event involving a near-extremal SMBH with a sufficiently high SNR of $\gtrsim$ several tens, inclusion of self-forces (at least in first order) might break this limitation and enable higher measurement precision for the SMBH spin.

\section*{Acknowlegement}
We thank Norichika Sago, Hiroyuki Nakano, Soichiro Morisaki, and Hayato Motohashi for fruitful discussions. We also thank Atsushi Nishizawa and Hiroki Takeda for valuable comments.
D. W. is supported by JSPS KAKENHI grant No. 23KJ06945.
N.O. is supported by the Grant-in-Aid for Scientific Research (KAKENHI) project (23K13111) and by the Hakubi project at Kyoto University.
D. T. is supported by the Sherman Fairchild Postdoctoral Fellowship at the California Institute of Technology.
\nocite{*}

\appendix
\section{Full noise curve for LISA}
\label{ap:LISA_noise}
The full noise curve of LISA $S(f)\:[\mathrm{Hz}{}^{-1}]$ in Eq.~\eqref{eq:inner_product_FD} is given by~\cite{LISA_noise_2019}
\begin{equation}
    S(f) = S_\mathrm{n}(f) + S_\mathrm{c}(f)\:,
\end{equation}
where $S_\mathrm{n}(f)$ is the instrumental noise and $S_\mathrm{c}(f)$ is the galactic confusion noise. $S_\mathrm{n}(f)$ is modeled by
\begin{gather}
    S_\mathrm{n}(f) = \frac{1}{\mathcal{F}(f)} \left( \frac{P_\mathrm{OMS}}{L^2} + (1+\cos^2(f/f_\ast))\frac{2P_\mathrm{acc}}{(2\pi f)^4 L^2} \right)\:,\\
    \mathcal{F}(f) = \frac{3}{10} (1+0.6(f/f_\ast)^2)^{-1}\:,
\end{gather}
where $2L=5\times 10^{11}$ cm is the round-trip travel distance of light ray, $f_\ast=c/(2\pi L)\approx 19.1$ mHz is the transfer frequency, $P_\mathrm{OMS}$ and $P_\mathrm{acc}$ are the single-link optical metrology noise and the single test mass acceleration noise, respectively. $\mathcal{F}(f)$ is the antenna pattern functions averaged over the sky position and polarization. $S_\mathrm{c}(f)$ is modeled by 
\begin{equation}
    S_\mathrm{c}(f) = Af^{-7/3} e^{-f^\alpha + \beta f\sin{(\kappa f)}} \left( 1+\tanh(\gamma(f_k-f)) \right)\:,
\end{equation}
where $A=9\times 10^{-45}$ and the parameters $\{ \alpha, \beta, \gamma, \kappa, f_k \}$ are fixed with the values for a four-year mission~\cite{LISA_noise_2019}. 

\section{Insensitivity of the measurement error on the choice of $t_\ast$}
\begin{table*}[h]
\begin{ruledtabular}
\caption{\label{tb:error_005}%
Same as Table.~\ref{tb:error_1}, but $t_\ast$ is set to the earliest time at which $\mathcal{M}_{lm} \leq 0.05$ for all modes.}
\begin{tabular}{cccccccc}
 BH spin & SNR of $(2,2)$ & SNR of $(3,3)$ & SNR of $(4,4)$ & SNR of $(5,5)$ & $\delta M/M [\%]$ & $\delta j/j [\%]$ & $c_{\mathrm{log}{M}\mathrm{log}j}$ \\
\colrule
$j=0.99$ & 27.5 & 31.6 & 55.9 & 56.4 & $0.222$ & $0.0384$ & $-0.985$ \\
$j=0.95$ & 56.8 &
89.0 & 52.2 & 21.4
 & $0.289$ & $0.127$ & $-0.970$ \\
$j=0.9$ & 86.3 &
69.5 & 32.4 & 14.1 & $0.464$ & $0.324$ & $-0.967$ \\
$j=0.8$ & 95.2 &
45.9 & 
21.1 & 9.79 & $0.902$ & 1.12 & $-0.960$
\end{tabular}
\end{ruledtabular}
\end{table*}

\begin{table*}[h]
\begin{ruledtabular}
\caption{\label{tb:error_01}%
Same as Table.~\ref{tb:error_1}, but $t_\ast$ is set to the earliest time at which $\mathcal{M}_{lm} \leq 0.1$ for all modes.}
\begin{tabular}{cccccccc}
 BH spin & SNR of $(2,2)$ & SNR of $(3,3)$ & SNR of $(4,4)$ & SNR of $(5,5)$ & $\delta M/M [\%]$ & $\delta j/j [\%]$ & $c_{\mathrm{log}{M}\mathrm{log}j}$ \\
\colrule
$j=0.99$ & 31.7 & 36.9 & 65.0 & 58.9 & $0.178$ & $0.0307$ & $-0.981$ \\
$j=0.95$ & 83.2 &
94.0 & 56.7 & 22.7
 & $0.257$ & $0.116$ & $-0.969$ \\
$j=0.9$ & 108 &
77.5 & 30.7 & 13.7 & $0.424$ & $0.296$ & $-0.964$ \\
$j=0.8$ & 110 &
47.1 & 
21.9 & 9.69 & $0.883$ & 1.07 & $-0.962$
\end{tabular}
\end{ruledtabular}
\end{table*}
\label{ap:insensitivity}
Since a large mismatch might lead to a misidentification of the source properties in the GW data analysis, we check how the choice of the fit time $t_\ast$ impacts the error estimates. Table~\ref{tb:error_005} and \ref{tb:error_01} show SNRs, the errors, and the correlation coefficient evaluated by fixing $t_\ast$ at the earliest time at which $\mathcal{M}_{lm}$ for all modes is less than or equal to $0.05$ and $0.1$ with $L_\mathrm{p}=1$, respectively. The expected measurement errors of $M$ and $j$ are found to be robust, affected only by factor of $\lesssim 2$,  against the change of the upper bound of the mismatch, i.e., $\mathcal{M}_{lm}=0.01$ and $\mathcal{M}_{lm}=0.1$.

\section{Additional Figures}
\label{ap:figures}
\begin{figure*}[h]
\centering
\includegraphics[scale=0.47]{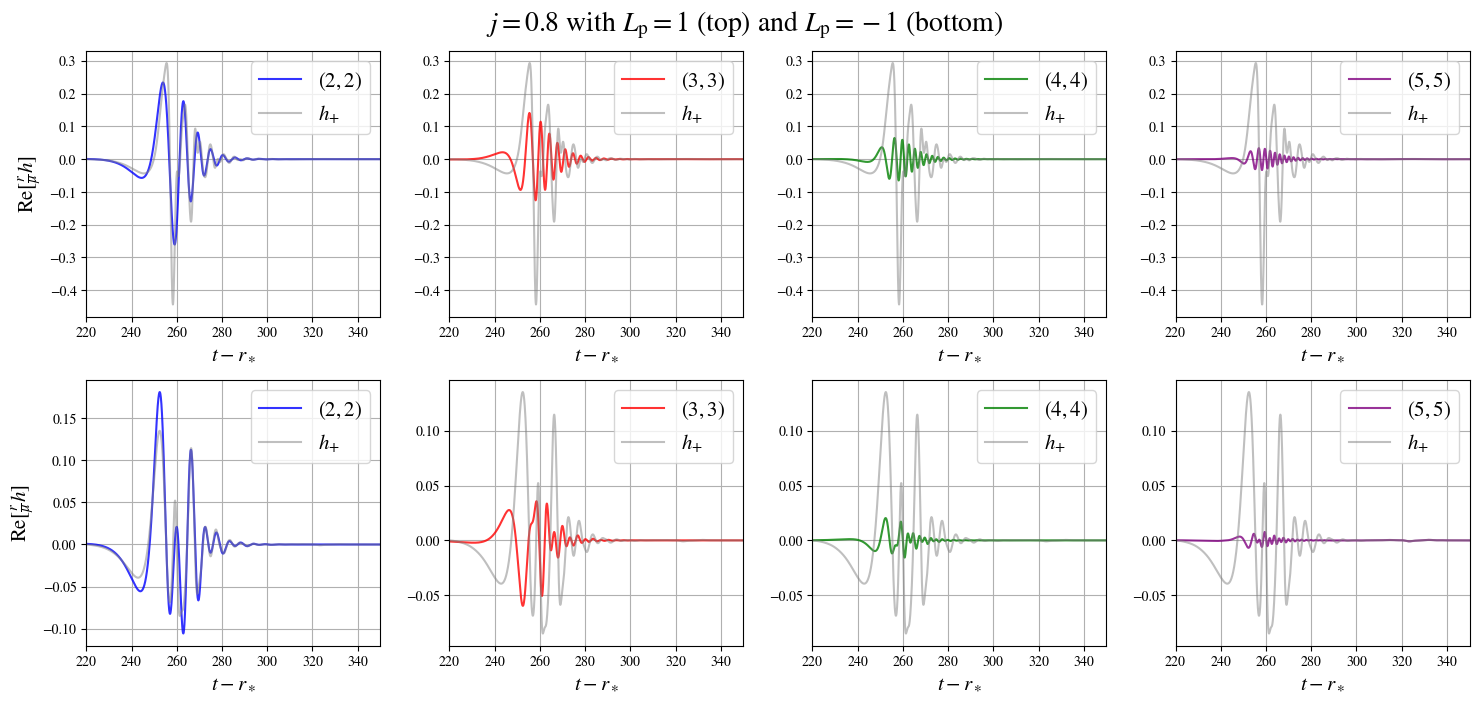}
\caption{Same as Fig.~\ref{fig:gws_099}, but for a SMBH spin of $j=0.8$.}
\label{fig:gws_08}
\end{figure*}

\begin{figure*}[h]
\centering
\includegraphics[scale=0.47]{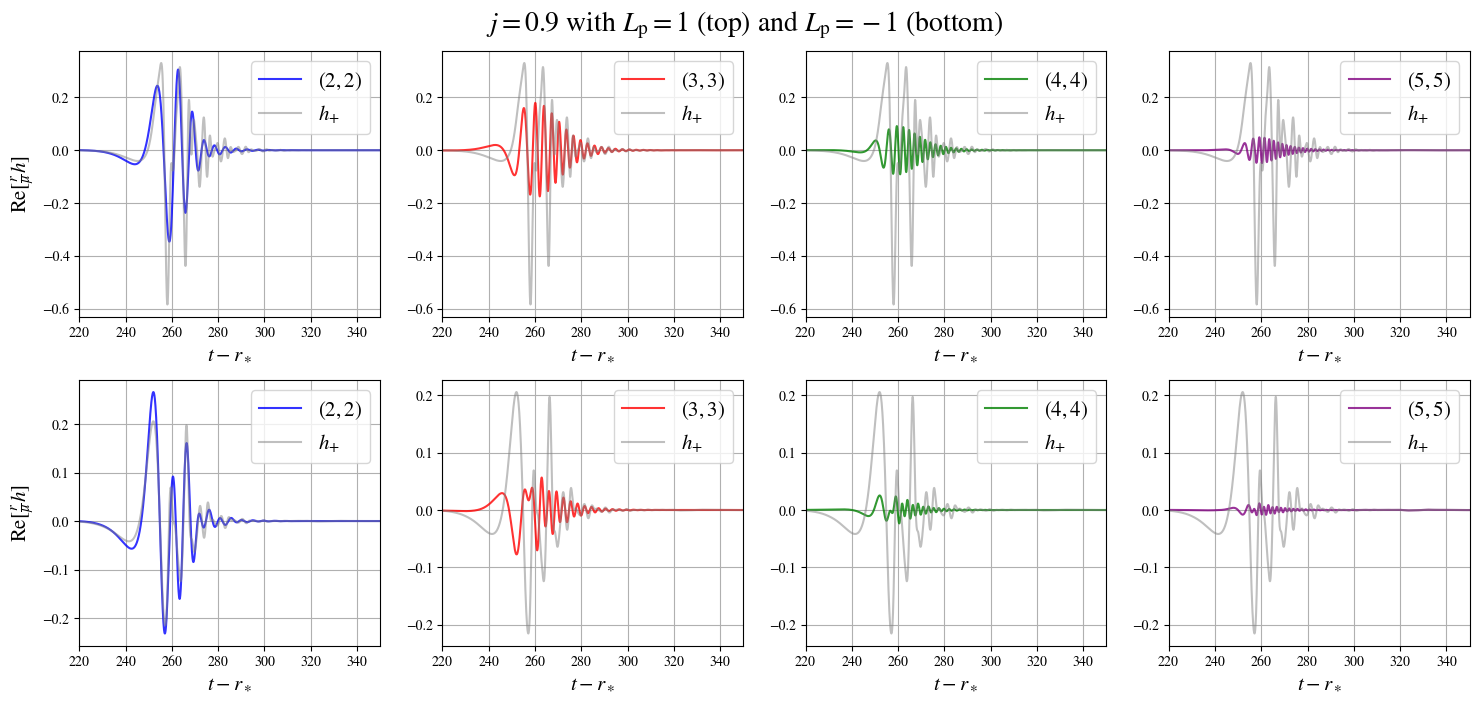}
\caption{Same as Fig.~\ref{fig:gws_099}, but for a SMBH spin of $j=0.9$.}
\label{fig:gws_09}
\end{figure*}

\begin{figure*}[h]
\centering
\includegraphics[scale=0.47]{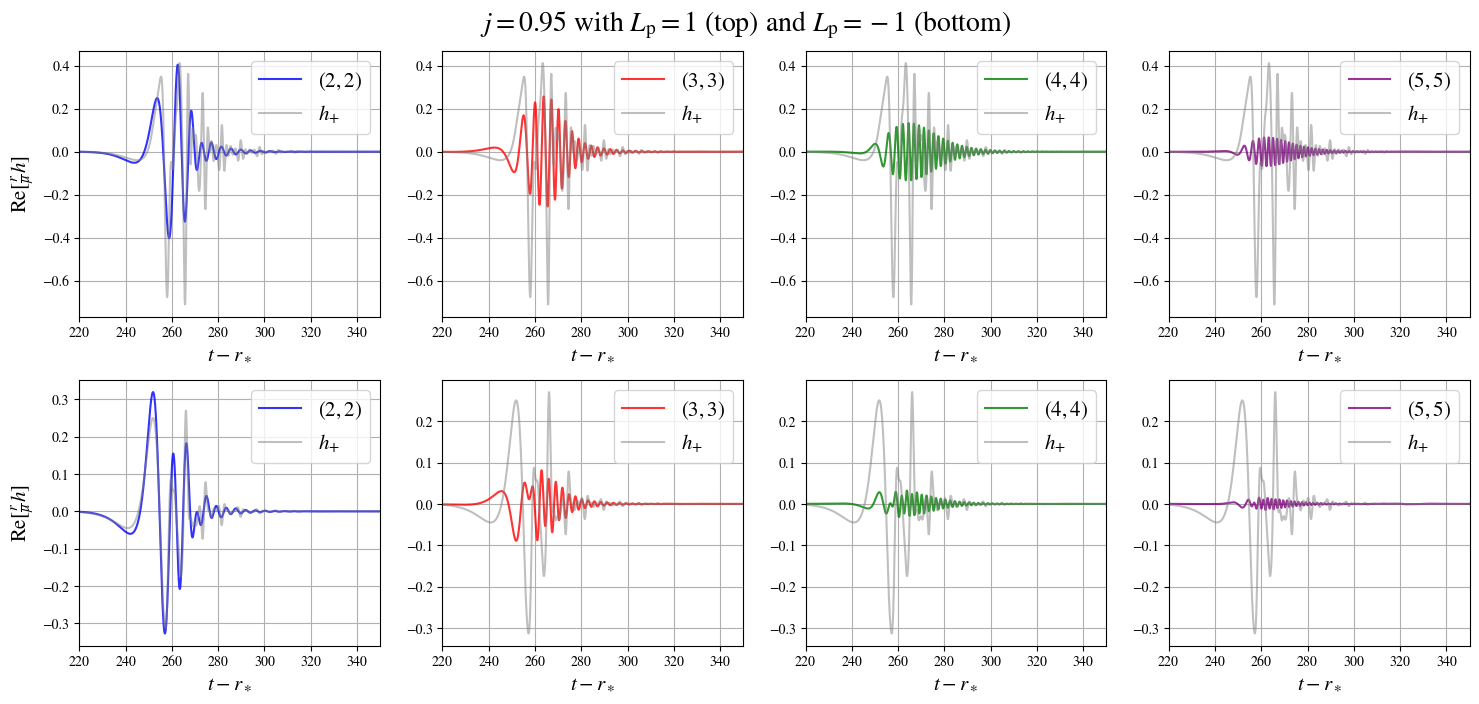}
\caption{Same as Fig.~\ref{fig:gws_099}, but for a SMBH spin of $j=0.95$.}
\label{fig:gws_095}
\end{figure*}

\begin{figure*}[h]
\centering
\includegraphics[scale=0.38]{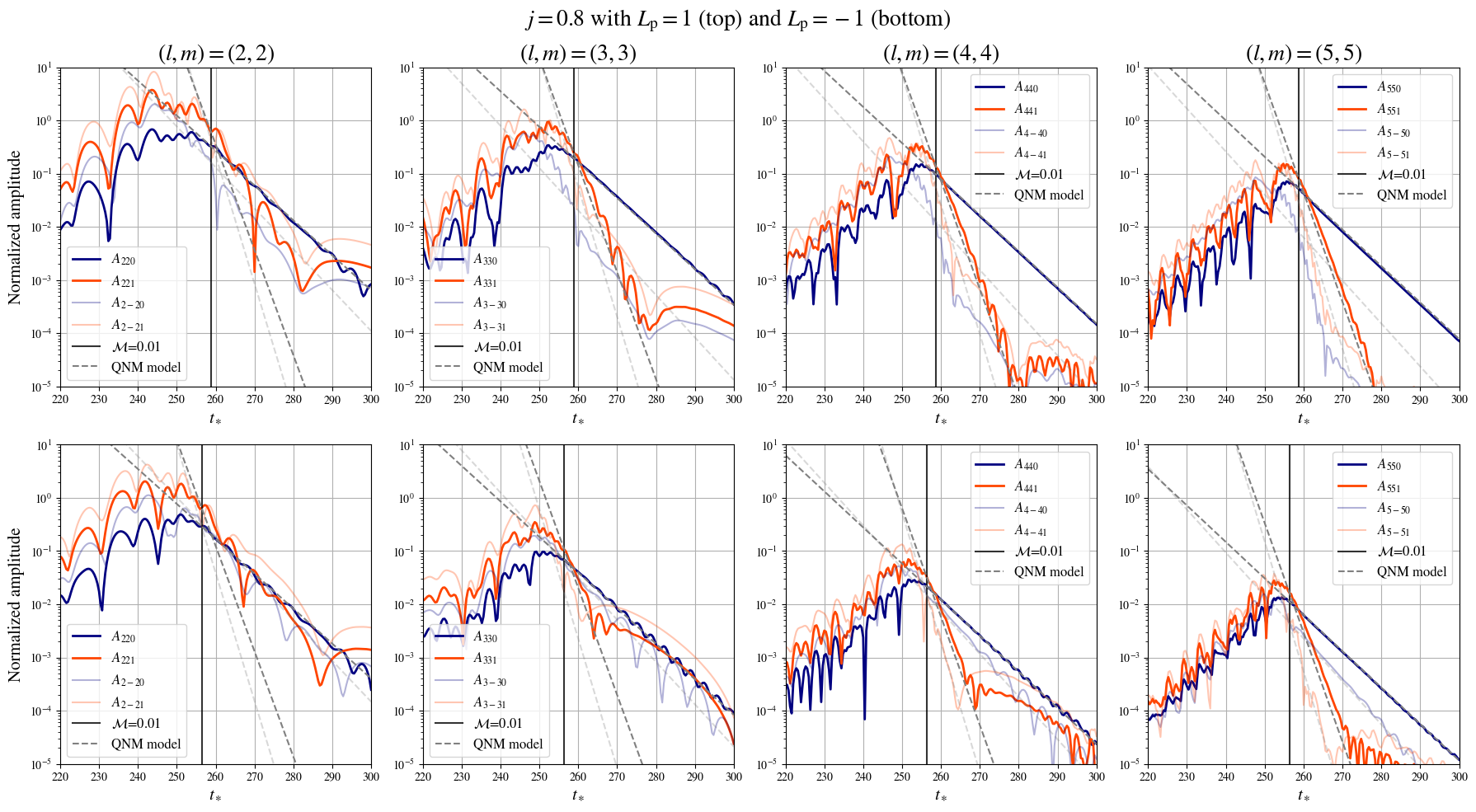}
\caption{Same as Fig.~\ref{fig:amp_099}, but for a SMBH spin of $j=0.8$.}
\label{fig:amp_08}
\end{figure*}

\begin{figure*}[h]
\centering
\includegraphics[scale=0.38]{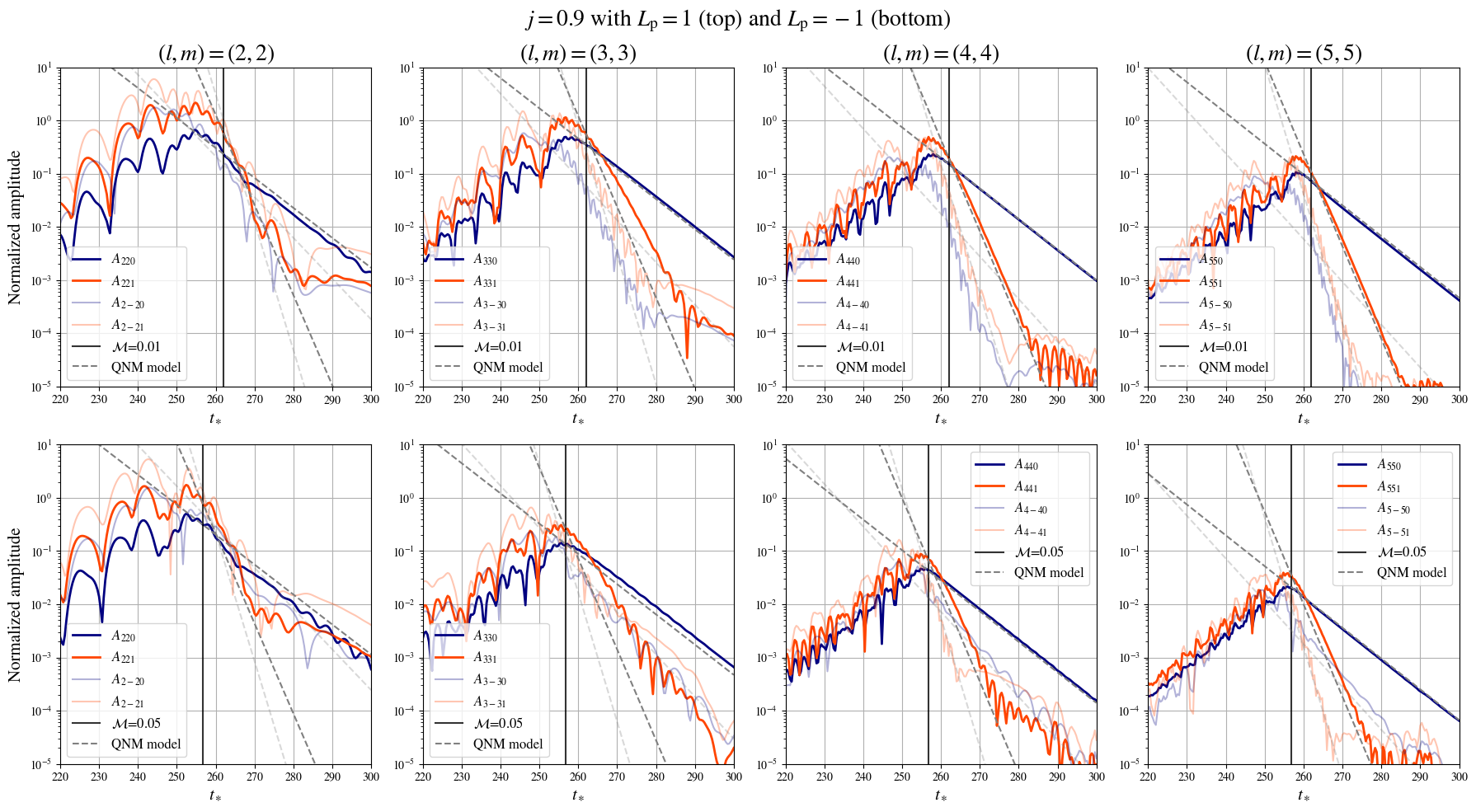}
\caption{Same as Fig.~\ref{fig:amp_099}, but for a SMBH spin of $j=0.9$.}
\label{fig:amp_09}
\end{figure*}

\begin{figure*}[h]
\centering
\includegraphics[scale=0.38]{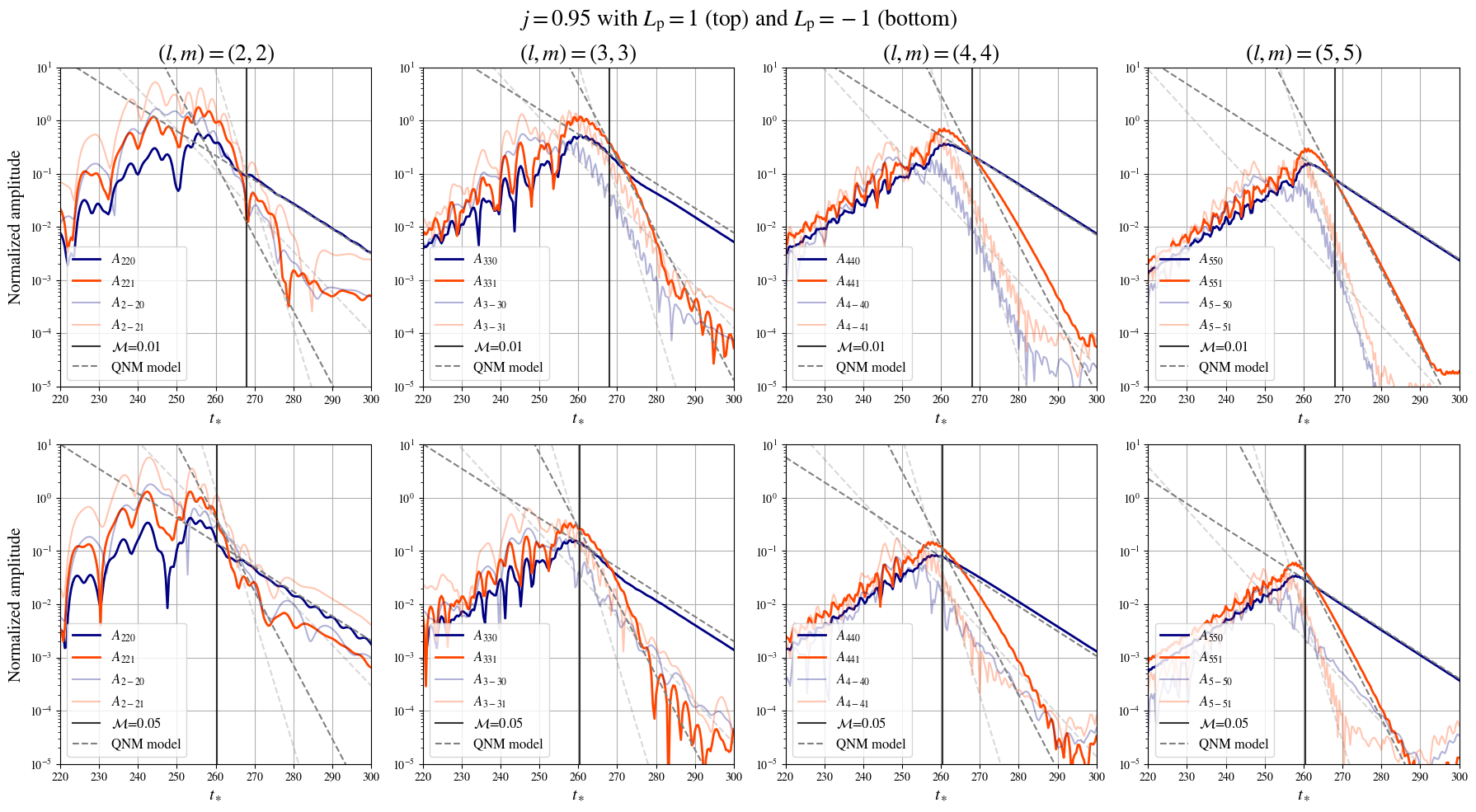}
\caption{Same as Fig.~\ref{fig:amp_099}, but for a SMBH spin of $j=0.95$.}
\label{fig:amp_095}
\end{figure*}

In the main text we have presented the waveforms and amplitude fits only for $j=0.99$. Here, we show our detailed results for the other spins: numerical waveforms we obtained by solving the Sasaki-Nakamura equation in Figs.~\ref{fig:gws_08}, \ref{fig:gws_09}, and \ref{fig:gws_095} and the best-fit amplitude for each QNM included in our QNM model in Figs.~\ref{fig:amp_08}, \ref{fig:amp_09}, and \ref{fig:amp_095}.


\newpage
\bibliographystyle{apsrev4-2}
\bibliography{bibfiles/reference}

\end{document}